\documentstyle[preprint,tighten,pra,aps]{revtex}
\begin{document}
\draft


\title{\LARGE \bf Relativistic Wigner Function, Charge Variable and Structure of Position Operator
\footnote{Contribution to the Seventh International Conference on
Squeezed States and Uncertainty Relations, Boston, Massachusetts,
USA, 4--7 June 2001}}

\author{B I Lev \footnote{%
Electronic mail: lev@iop.kiev.ua}, A A Semenov \footnote{%
Electronic mail: sem@iop.kiev.ua}}
\address{Institute of Physics,
National Academy of Science of Ukraine, 46 Nauky pr, Kiev 03028,
Ukraine}
\author{ C V Usenko \footnote{%
Electronic mail: usenko@ups.kiev.ua}}
\address{ Physics
Department, Taras Shevchenko Kiev University, 6 Academician
Glushkov pr, Kiev 03127, Ukraine}

\maketitle

\begin{abstract}

The relativistic phase-space representation by means of the usual
position and momentum operators for a class of observables
with Weyl symbols independent of charge variable (i.e. with any
combination of position and momentum) is proposed. The dynamical
equation coincides with its analogue in the non-local theory
(generalization of the Newton-Wigner position operator approach)
under conditions when particles creation is impossible.
Differences reveal themselves in specific constraints on possible
initial conditions.
\end{abstract}


\section{Introduction}
Attempts to generalize the Weyl-Wigner-Moyal (WWM) formalism for
relativistic case lead to a number of problems. The first problem
is that Weyl rule does not include time as a dynamical variable,
and the scalar product in Hilbert space of states is formulated
not for functions square integrable over the whole space-time but
in the three-dimensional space or on a space-like hyper-surface
only. It leads to appearance of non Lorentz invariant formulation
for the star-product and, as a consequence, to non Lorentz
invariant quantum Liouville equation as well. We suppose that in
the contemporary approaches this problem can be resolved in the
framework of Tomonaga - Schwinger description of quantum field
theory \cite{b1,b2}.

Next essential problem in relativistic WWM formalism is absence of
a well-defined position operator. Indeed, even for a free particle
the usual position operator is not one-particle one. Furthermore,
there are some problems with semi-classical limit here. They could
be resolved by introduction of one-particle well defined Newton -
Wigner position operator \cite{b3}, however one meets difficulties
related to Lorentz invariance as well, but at the level of the
wave equation. We propose formulation of the WWM formalism by
means of the usual position operator but for such class of
observables that their Weyl symbols are independent of the charge
variable (we denominate them as the charge-invariant observables).
It is possible to introduce the usual (not matrix-valued) Wigner
function here 
. This object includes four components: two correspond to particle
and antiparticle (even part of Wigner function), and two more are
interference terms (odd part of Wigner function, that is to make
no effect on expected values of observables because of the charge
superselection rule ). Evolution equation (in the case when
one-particle interpretation of relativistic quantum mechanics is
possible, i.e. for a free particle and a particle in a static
magnetic field) for odd part turns it in classical limit to zero,
and equation for even part coincides with its analogue in the
Newton - Wigner position operator \cite{b4} and in the non-local
theory \cite{b41,b42} approaches. The difference reveals itself in
peculiarities of constraints on initial conditions \cite{b5}.

We consider how expected values of the charge-invariant
observables distinguish between in two approaches to definition of
position operator. We find it very important, all the more that
there are suppositions that it can be related to the existence of
the preferred frame in the Universe \cite{b6}. In that case its
introduction is related to attempts of correct tachyons
description and, as a consequence, to a possible explanation of
instant quantum correlations in relativistic case.

\section{Symbols, definitions and representation}
Here we use the Feshbach-Villars formalism for the Klein - Gordon
equation \cite{b61}. If one applies the standard change of wave function
\begin{equation}
\begin{tabular}{l}
 $\psi  = \frac{1}{{\sqrt 2 }}(\varphi  + \chi )$ \\
 $i\hbar \frac{{\partial \psi }}{{\partial t}} = \frac{{mc^2 }}{{\sqrt 2 }}(\varphi  - \chi
 )$, \label{f1}
\end{tabular}
\end{equation}
this equations can be written in the form of the usual Sr\"odinger
equation for two-component wave function with the Hamiltonian:
\begin{equation}
\hat H = (\tau _3  + i\tau _2 )\frac{{(\hat{\vec p} - e\vec
A(x))^2 }}{{2m}} + \tau _3 mc^2  + e\varphi (x). \label{f2}
\end{equation}

The transform matrix to the Feshbach-Villars (energy)
representation has the following form:
\begin{equation}
\begin{tabular}{l}
$U(\hat p)=\frac{1}{{2\sqrt{mc^2 E(\hat p)}}}\left[
{(E(\hat p)+mc^2)+(E(\hat p)-mc^2)\tau _1 }\right]$\\
$U(\hat n) = \frac{1}{{2\sqrt {mc^2 E(\hat n)} }}\left[ {(E(\hat
n) + mc^2 ) + (E(\hat n) - mc^2 )\tau _1 } \right]$. \label{f3}
\end{tabular}
\end{equation}
Where $E(p)$ and $E(n)$ are the expressions for the energy of a
free particle and the spectrum of relativistic rotator:
\begin{equation}
\begin{tabular}{l}
$E(p) = \sqrt {m^2 c^4  + c^2 p^2 }$ \\ $E(n) = \sqrt {m^2 c^4  +
2mc^2 \hbar \omega (n + \frac{1}{2})}$ . \label{f4}
\end{tabular}
\end{equation}
Following combination of the matrix (\ref{f3}) is frequently used:
\begin{equation}
R(s_1 ,s_2 ) = U(s_1 )U^{ - 1} (s_2 ) = \varepsilon (s_1 ,s_2 ) +
\chi (s_1 ,s_2 )\tau _1 , \label{f5}
\end{equation}
here $s = p,n$. It contains even and odd parts and is expressed
via the energy (\ref{f4}):
\begin{equation}
\begin{tabular}{l}
$\varepsilon (s_1 ,s_2 ) = \frac{{E(s_1 ) + E(s_2 )}}{{2\sqrt
{E(s_1 )E(s_2 )} }}$\\ $\chi (s_1 ,s_2 ) = \frac{{E(s_1 ) - E(s_2
)}}{{2\sqrt {E(s_1 )E(s_2 )} }}$. \label{f6}
\end{tabular}
\end{equation}

\section{Matrix-valued phase-space representation for scalar charged particles}
The matrix-valued Weyl transform of an operator $\hat
A_\alpha{}^\beta$ is determined in the following way:
\begin{equation}
\hat A_\alpha {}^\beta   = \sum\limits_{\gamma  =  \pm 1}
{\int\limits_{ - \infty }^{ + \infty } {A_\gamma  {}^\beta
(p,q)\hat W_\alpha  {}^\gamma  (p,q)dpdq} }. \label{f8}
\end{equation}
$\hat W_\alpha  {}^\beta  (p,q)$ is the operator of
quasi-probability density:
\begin{equation}
\hat W_\alpha  {}^\beta  (p,q) = \frac{1}{{(2\pi \hbar )^d
}}\int\limits_{ - \infty }^{ + \infty } {\left| {q + \frac{Q}{2}}
\right\rangle \delta _\alpha  {}^\beta  e^{\frac{i}{\hbar }Qp}
dQ\left\langle {q - \frac{Q}{2}} \right|}. \label{f9}
\end{equation}
Inverse transformation has the following form:
\begin{equation}
A_\alpha  {}^\beta  (p,q) = \int\limits_{ - \infty }^{ + \infty }
{\left\langle {q + \frac{Q}{2}} \right|\hat A_\alpha  {}^\beta
\left| {q - \frac{Q}{2}} \right\rangle e^{ - \frac{i}{\hbar }Qp}
dQ}. \label{f10}
\end{equation}

The matrix-valued Moyal bracket can be determined in the
following form:
\begin{equation}
\left\{\check{A}(p,q),\check{B}(p,q)\right\}_M  =
\frac{1}{{i\hbar }}\left( \check{A}(p,q) \star \check{B}(p,q) -
\check{B}(p,q) \star \check{A}(p,q)\right).\label{f11}
\end{equation}

The matrix-valued Weyl transformation of the density matrix leads
to the matrix-valued Wigner function:
\begin{equation}
W_\alpha {}^\beta  (p,q) = \frac{1}{{(2\pi \hbar )^d
}}\int\limits_{ - \infty }^{ + \infty } {\psi _\alpha ^ *  (q +
\frac{Q}{2})\psi ^\beta  (q - \frac{Q}{2})e^{\frac{i}{\hbar }Qp}
dQ}.\label{f12}
\end{equation}

After standard transformations, similar to the usual WWM
formalism, we obtain the quantum Liouville equation
\begin{equation}
\partial _t \check{W} = \left\{ \check{H},\check{W}\right\} _M .\label{f13}
\end{equation}

{\bf Why this formalism is not Lorentz invariant?} Possible
explanation in the framework of the Tomonaga-Schwinger approach
to quantum field theory.

The fact is that average values calculated in this approach
coincide with ones in the usual (Schr\"odinger) representation of
quantum mechanics that is Lorentz invariant. Nevertheless, the
scalar product is determined here with functions that are square
integrable in a certain space-like hyper-surface. One can relate
this hyper-surface to the measuring device frame. In other words,
the wavefunction collapse occurs in a frame in which the equation
(\ref{f13}) is written. The absence of Lorentz invariance is a
consequence of the fact that the Weyl rule does not include time
as an independent dynamical variable. Note, that
equation (\ref{f13}) can be written with four-dimensional Lorentz
invariant symbols only, but to do this we have to involve into
consideration a certain time-like unit vector in a way similar to the
Tomonaga-Schwinger approach to quantum field theory. It is the
four-velocity of the frame where wavefunction reduction occurs
(the measuring device frame) relative to the second static
observer (watching observer). In our approach this process obeys
the relativity of simultaneity. As a result, at a certain
instant, in a static frame (attached to the watching observer), a
state with a part of the wavefunction before measurement on one
hand and a part of the wave function after measurement on another
hand, is realized.

Hence we find to be very important and of fundamental value the
fact that quantum mechanics in the Wigner representation
necessarily includes the four-velocity of the measuring device
frame in explicit form.

\section{Charge-invariant observables in the Feshbach-Villars representation. Free particle}
Let the matrix-valued Weyl symbol be proportional to the identity
matrix:
\begin{equation}
A_\alpha {}^\beta  (p,q) = A(p,q)\delta _\alpha {}^\beta
.\label{f19}
\end{equation}
Such symbols do not depend on the
charge variable, so that the class of dynamical variables, which
corresponds to these we denominate here, are more brief, as a
class of charge-invariant observables. Most of dynamical
variables that we consider in relativistic (non-quantum)
mechanics belong to this class. The reason for this is the
absence of a dependence on the charge variable in classical
mechanics.

The Weyl transformation for charge-invariant observables in the
Feshbach-Villars representation has the form:

\begin{equation}
\hat A^{FV}{}_\alpha {}^\beta   = \frac{1}{{(2\pi \hbar )^d
}}\int\limits_{ - \infty }^{ + \infty } {\left| {p + \frac{P}{2}}
\right\rangle A(p,q)R_\alpha {}^\beta  (p + \frac{P}{2},p -
\frac{P}{2})e^{ - \frac{i}{\hbar }Pq} dPdpdq\left\langle {p -
\frac{P}{2}} \right|}.\label{f20}
\end{equation}
Unlike the Newton-Wigner coordinate approach and the
non-relativistic case there is a matrix-valued function (\ref{f5})
here.

Consequences from (\ref{f20}) are expressions for even $[\hat A]$
and odd $\{ \hat A\}$ parts of the operator of a charge-invariant
observable in terms of its Weyl symbol:
\begin{equation}
[\hat A^{FV} ]_\alpha  {}^\beta   = \frac{1}{{(2\pi \hbar )^d
}}\int\limits_{ - \infty }^{ + \infty } {\left| {p + \frac{P}{2}}
\right\rangle A(p,q)\varepsilon (p + \frac{P}{2},p -
\frac{P}{2})\delta _\alpha  {}^\beta  e^{ - \frac{i}{\hbar }Pq}
dPdpdq\left\langle {p - \frac{P}{2}} \right|} ,\label{f21}
\end{equation}
\begin{equation}
\{ \hat A^{FV} \} _\alpha {} ^\beta   = \frac{1}{{(2\pi \hbar )^d
}}\int\limits_{ - \infty }^{ + \infty } {\left| {p + \frac{P}{2}}
\right\rangle A(p,q)\chi (p + \frac{P}{2},p - \frac{P}{2})\tau
_1{}_\alpha {}^\beta  e^{ - \frac{i}{\hbar }Pq}
dPdpdq\left\langle {p - \frac{P}{2}} \right|}. \label{f22}
\end{equation}

Then, one can obtain a formula that reconstructs the Weyl symbol
of operator, with even and odd parts, in the Feshbach - Villars
representation:
\begin{equation}
A(p,q)\delta _\alpha {}^\beta   = \sum\limits_{\gamma  =  \pm 1}
{\int\limits_{ - \infty }^{ + \infty } {R^{ - 1}{} _\gamma
{}^\beta (p - \frac{P}{2},p + \frac{P}{2})\left\langle {p +
\frac{P}{2}} \right|\hat A^{FV} {}_\alpha {} ^\gamma  \left| {p -
\frac{P}{2}} \right\rangle e^{\frac{i}{\hbar }Pq} dP}
},\label{f23}
\end{equation}
\begin{equation}
A(p,q)\delta _\alpha  {}^\beta   = \int\limits_{ - \infty }^{ +
\infty } {\varepsilon ^{ - 1} (p - \frac{P}{2},p +
\frac{P}{2})\left\langle {p + \frac{P}{2}} \right|[\hat A^{FV}
]_\alpha {}^\beta  \left| {p - \frac{P}{2}} \right\rangle
e^{\frac{i}{\hbar }Pq} dP},\label{f24}
\end{equation}
\begin{equation}
A(p,q)\delta _\alpha  {}^\beta   = \sum\limits_{\gamma  =  \pm 1}
{\int\limits_{ - \infty }^{ + \infty } {\chi ^{ - 1} (p -
\frac{P}{2},p + \frac{P}{2})\tau _1 {}_\gamma  {}^\beta
\left\langle {p + \frac{P}{2}} \right|\{ \hat A^{FV} \} _\alpha
{}^\gamma \left| {p - \frac{P}{2}} \right\rangle
e^{\frac{i}{\hbar }Pq} dP} }.\label{f25}
\end{equation}

Comparing (\ref{f24}) and (\ref{f25}) we conclude that matrix
elements (integral kernels) of even and odd parts of the operator
of an arbitrary charge-invariant variable are uniquely related to
each other due to the Weyl rule:
\begin{equation}
\left\langle {p_1 } \right|\{ \hat A^{FV} \} _\alpha  {}^\beta
\left| {p_2 } \right\rangle  = \frac{{E(p_1 ) - E(p_2 )}}{{E(p_1
) + E(p_2 )}}\sum\limits_{\gamma  =  \pm 1} {\tau _1 {}_\gamma
{}^\beta  \left\langle {p_1 } \right|[\hat A^{FV} ]_\alpha
{}^\gamma \left| {p_2 } \right\rangle }.\label{f26}
\end{equation}

A consequence from this expression is the fact that odd part of
an operator independent on position is zero. If one uses as $\hat
A$ , for example, the scalar potential of an electric field,
expression (\ref{f26}) establishes a quantitative relationship
between the effects of motion of a particle in an electric field
and its interaction with polarizable vacuum (trembling motion,
Zitterbewegung).

\section{Charge-invariant observables in the Feshbach-Villars representation. Constant magnetic field}
We will provide a similar consideration for a particle in a
constant magnetic field in the energy representation. We
introduce a hermitian generalization of the Wigner function:
\begin{equation}
W_{nm} (p,q) = \frac{1}{{(2\pi \hbar )^d }}\int\limits_{ - \infty
}^{ + \infty } {\varphi _m^ *  \left( {p + \frac{P}{2}}
\right)\varphi _n \left( {p - \frac{P}{2}} \right)e^{ -
\frac{i}{\hbar }Pq} dP},\label{f27}
\end{equation}
where $\varphi _n (p)$ is the momentum part of the eigenfunction
of the Hamiltonian.

The Weyl transformation for charge-invariant observables in the
energy representation has the form:

\begin{equation}
A^E {}_{nm} {}_\alpha  {}^\beta   = R_\alpha  {}^\beta
(m,n)\int\limits_{ - \infty }^{ + \infty } {A(p,q)W_{nm}
(p,q)dpdq}. \label{f28}
\end{equation}

Unlike the non-local theory and the non-relativistic case there
is a matrix-valued function (\ref{f5}) here.

Consequences from (\ref{f28}) are expressions for even $[\hat A]$
and odd $\{ \hat A\}$ parts of the operator of a charge-invariant
observable in terms of its Weyl symbol:
\begin{equation}
\left[ {A^E } \right]_{nm} {}_\alpha {}^\beta   = \varepsilon
(m,n)\delta _\alpha  {}^\beta  \int\limits_{ - \infty }^{ + \infty
} {A(p,q)W_{nm} (p,q)dpdq},\label{f29}
\end{equation}
\begin{equation}
\left\{ {A^E } \right\}_{nm} {}_\alpha {}^\beta   = \chi
(m,n)\tau _1 {}_\alpha  {}^\beta  \int\limits_{ - \infty }^{ +
\infty } {A(p,q)W_{nm} (p,q)dpdq},\label{f30}
\end{equation}

Then, one can obtain a formula that reconstructs the Weyl symbol
of operator, with even and odd parts, in the energy
representation:
\begin{equation}
A(p,q)\delta _\alpha {}^\beta   = \sum\limits_{\scriptstyle
\gamma  =  \pm 1 \hfill \atop
  \scriptstyle mn \hfill
  }{R^{ - 1} {}_\gamma  {}^\beta  (m,n)A^E {}_{nm} {}_\alpha {}^\gamma  W_{mn} (p,q)}, \label{f31}
\end{equation}
\begin{equation}
A(p,q)\delta _\alpha  {}^\beta   = \sum\limits_{mn} {\varepsilon
^{ - 1} (m,n)\left[ {A^E } \right]_{nm} {}_\alpha  {}^\beta
W_{mn} (p,q)},\label{f32}
\end{equation}
\begin{equation}
A(p,q)\delta _\alpha  {}^\beta   = \sum\limits_{\scriptstyle
\gamma  =  \pm 1 \hfill \atop \scriptstyle mn \hfill }{\chi ^{ -
1} (m,n)\tau _1 {}_\gamma  {}^\beta \left\{ {A^E } \right\}_{nm}
{}_\alpha {}^\gamma  W_{mn}
  (p,q)}.\label{f33}
\end{equation}
Comparing (\ref{f32}) and (\ref{f33}) we conclude that matrix
elements (integral kernels) of even and odd parts of the operator
of an arbitrary charge-invariant variable are uniquely related to
each other due to the Weyl rule:
\begin{equation}
\{ \hat A^{FV} \} _{nm} {}_\alpha  {}^\beta   = \frac{{E(m) -
E(n)}}{{E(m) + E(n)}}\sum\limits_{\gamma  =  \pm 1} {\tau _1
{}_\gamma  {}^\beta  [\hat A^{FV} ]_{nm}{}_\alpha  {}^\gamma
}.\label{f34}
\end{equation}

\section{Wigner function and quantum Liouville equation for charge-invariant observables. Free particle}
It is easy to see from (9) that it is possible to introduce the
usual Wigner function for the charge-invariant observables in
such a way that their average values are determined by the
formula:
\begin{equation}
\bar A = \int\limits_{ - \infty }^{ + \infty } {A(p,q)W(p,q)dpdq}
.\label{f35}
\end{equation}
The Wigner function can be determined as the sum of the four
following components:
\begin{equation}
W_\alpha  {}^\beta  (p,q) = \left\langle {\psi _\beta  }
\right|\hat W^{FV} {}_\alpha  {}^\beta  (p,q)\left| {\psi ^\alpha
} \right\rangle.\label{f36}
\end{equation}
And, in a fact, is the average value of the operator of
quasi-probability density in the Feshbach-Villars representation:
\begin{equation}
\hat W^{FV} {}_\alpha  {}^\beta  (p,q) = \frac{1}{{(2\pi \hbar )^d
}}\int {\left| {p + \frac{P}{2}} \right\rangle R_\alpha  {}^\beta
(p + \frac{P}{2},p - \frac{P}{2})e^{ - \frac{i}{\hbar }Pq}
dP\left\langle {p - \frac{P}{2}} \right|}.\label{f37}
\end{equation}
Substituting (\ref{f37}) into (\ref{f36}), we obtain for the
Wigner function components the following expressions:

\begin{equation}
W_\alpha  {}^\alpha  (p,q) = \frac{1}{{(2\pi \hbar )^d }}\int
{\varepsilon (p + \frac{P}{2},p - \frac{P}{2})\psi _\alpha ^ * (p
+ \frac{P}{2})\psi ^\alpha  (p - \frac{P}{2})e^{ - \frac{i}{\hbar
}Pq} dP}, \label{f38}
\end{equation}
\begin{equation}
W_\alpha  {}^{ - \alpha } (p,q) = \frac{1}{{(2\pi \hbar )^d }}\int
{\chi (p + \frac{P}{2},p - \frac{P}{2})\psi _\alpha ^ *  (p +
\frac{P}{2})\psi ^{ - \alpha } (p - \frac{P}{2})e^{ -
\frac{i}{\hbar }Pq} dP}. \label{f39}
\end{equation}

Even components of the Wigner function (\ref{f38}) correspond to a
charge definite state. The
value of odd components (\ref{f39}) for such a state is zero. The
expression (\ref{f38}) differs from analogous one for a
non-relativistic Wigner func-tion and relativistic one determined
using the Newton - Wigner position operator by the function
$\varepsilon (p_1 ,p_2 )$ under the integral sign (see
(\ref{f6})).

The following equations can be obtained in a standard way,
through differentiating Wigner function components with respect
to time:
\begin{equation}
\partial _t W_\alpha  {}^\alpha  (p,q,t) = \alpha \frac{2}{\hbar }E(p)\sin \left\{ -\frac{\hbar }{2}\left(
\overleftarrow {\partial } _p \overrightarrow {\partial }
_q\right)\right\} W_\alpha  {}^\alpha (p,q,t), \label{f40}
\end{equation}
\begin{equation}
\partial _t W_\alpha  {}^{ - \alpha } (p,q,t) = i\alpha \frac{2}{\hbar }E(p)\cos \left\{ \frac{\hbar }{2}\left(
\overleftarrow {\partial } _p \overrightarrow {\partial }
_q\right)\right\} W_\alpha  {}^{ - \alpha } (p,q,t). \label{f41}
\end{equation}

Nevertheless, the Wigner function components are not independent,
i.e. specific constraints are imposed on solutions of the system
(\ref{f40}), (\ref{f41}):
\begin{equation}
\begin{tabular}{l}
$(E(p_1 ) - E(p_2 ))^2 \int\limits_{ - \infty }^{ + \infty } {W_
+  {}^ +  (\frac{1}{2}(p_1  + p_2 ),q_1 )W_ - {} ^ -
(\frac{1}{2}(p_1  + p_2 ),q_2 )e^{\frac{i}{\hbar }(q_1  + q_2
)(p_1  - p_2 )} dq_1 dq_2  = }$ \\
$= (E(p_1 ) + E(p_2 ))^2 \int\limits_{ - \infty }^{ + \infty } {W_
+  {}^ -  (\frac{1}{2}(p_1  + p_2 ),q_1 )W_ - {} ^ {+}
(\frac{1}{2}(p_1  + p_2 ),q_2 )e^{\frac{i}{\hbar }(q_1  + q_2
)(p_1  - p_2 )} dq_1 dq_2 }$
\end{tabular} , \label{f42}
\end{equation}
\begin{equation}
W^ *  {}_\alpha  {}^\alpha  (p,q) = W_\alpha  {}^\alpha  (p,q) ,
\label{f43}
\end{equation}
\begin{equation}
W^ *  {}_\alpha  ^{ - \alpha } (p,q) = W_{ - \alpha } {}^\alpha
(p,q).\label{f44}
\end{equation}

It is essential that the equation for even components of the
Wigner function coincides with the analogous expression obtained
for the formalism where the Newton-Wigner position operator is
used. Hence, the dynamics of quasi-distribution functions for
charge-definite or mixed on the charge variable states is
identical in both cases.

\section{Wigner function and quantum Liouville equation for charge-invariant observables. Constant magnetic field}
The Wigner function components in the case of a particle in a
constant magnetic field can be determined in two ways: via the
wavefunction in the energy representation

\begin{equation}
W_\alpha  {}^\alpha  (p,q) = \sum\limits_{nm} {\varepsilon (m,n)}
W_{nm} (p,q)C_m^ * {} _\alpha  C_n{} ^\alpha, \label{f45}
\end{equation}
\begin{equation}
W_\alpha  {}^{-\alpha}  (p,q) = \sum\limits_{nm} {\chi (m,n)}
W_{nm} (p,q)C_m^ *  {}_\alpha  C_n {}^{ - \alpha }, \label{f46}
\end{equation}
and via the wavefunction in the representation of the non-local
theory
\begin{equation}
W_\alpha  {}^\alpha  (p,q) = \frac{1}{{(2\pi \hbar )^d }}\int
{\varepsilon (p + \frac{P}{2},p_1 ;p - \frac{P}{2},p_2 )\psi
_\alpha {}^ *  (p_1 )\psi ^\alpha  (p_2 )e^{ - \frac{i}{\hbar }Pq}
dPdp_1 dp_2 }, \label{f47}
\end{equation}
\begin{equation}
W_\alpha  ^{ - \alpha } (p,q) = \frac{1}{{(2\pi \hbar )^d }}\int
{\chi (p + \frac{P}{2},p_1 ;p - \frac{P}{2},p_2 )\psi _\alpha ^
*  (p_1 )\psi ^{ - \alpha } (p_2 )e^{ - \frac{i}{\hbar }Pq}
dPdp_1 dp_2 }.\label{f48}
\end{equation}
Here we have introduced the following generalized functions:
\begin{equation}
\varepsilon (p',p_1 ;p'',p_2 ) = \sum\limits_{nm} {\varepsilon
(m,n)\varphi _m^ *  (p')\varphi _m (p_1 )\varphi _n (p'')\varphi
_n^
*  (p_2 )}, \label{f49}
\end{equation}
\begin{equation}
\chi (p',p_1 ;p'',p_2 ) = \sum\limits_{nm} {\chi (m,n)\varphi _m^
* (p')\varphi _m (p_1 )\varphi _n (p'')\varphi _n^ *  (p_2
)}.\label{f50}
\end{equation}

The evolution equations can be obtained in a standard way,
through applying the well-known expressions for a Hermitian
generalization of the Wigner function:
\begin{equation}
\partial _t W_\alpha  {}^\alpha  (p,q,t) = \alpha \frac{2}{\hbar }E(p,q)\sin \left\{ \frac{\hbar }{2}\left(\overleftarrow {\partial } _q \overrightarrow {\partial } _p  -
\overleftarrow {\partial } _p \overrightarrow {\partial }
_q\right)\right\} W_\alpha  {}^\alpha (p,q,t), \label{f51}
\end{equation}
\begin{equation}
\partial _t W_\alpha  {}^{ - \alpha } (p,q,t) = i\alpha \frac{2}{\hbar }E(p,q)\cos \left\{ \frac{\hbar }{2}\left(\overleftarrow {\partial } _q \overrightarrow {\partial } _p  -
\overleftarrow {\partial } _p \overrightarrow {\partial }
_q\right)\right\}W_\alpha  {}^{ - \alpha } (p,q,t). \label{f52}
\end{equation}
Here we have introduced the following effective Hamiltonian:
\begin{equation}
E(p,q) = \sqrt[ \star ]{{m^2 c^4  + c^2 (p - eA(q))^2 }} .
\label{f53}
\end{equation}
It should be noticed that the square root is defined here by
means of the star-product $\star\equiv e^{\frac{i\hbar}{2}\left(\overleftarrow{\partial }_{q}%
\overrightarrow{\partial }_{p}-\overleftarrow{\partial }_{p}\overrightarrow{%
\partial }_{q}\right)}$. This is a common feature for the
both usual and non-local theories and it leads to the appearance
of specific peculiarities in a relativistic quantum particle
behavior, not related to the complicated charge structure of the
position operator.

Similar to the free particle case, one can find specific
constraints on solutions of the system (\ref{f51}), (\ref{f52}).
The expressions (\ref{f43}) and (\ref{f44}) are left without
modification, and the analog of the expression (\ref{f42}) has the
following form:
\begin{equation}
\begin{tabular}{l}
$\int\limits_{ - \infty }^{ + \infty } {\varepsilon ^{ - 1} (p_1
,p';p_2 ,p'')W_ + {} ^ +  (\frac{1}{2}(p_1  + p_2 ),q_1
)e^{\frac{i}{\hbar }(p_1  - p_2 )q} dp_1 dp_2 }
dq$\\
$\times\int\limits_{ - \infty }^{ + \infty } {\varepsilon ^{ - 1}
(p_1 ,p';p_2 ,p'')W_ -  {}^ -  (\frac{1}{2}(p_1  + p_2 ),q_2
)e^{\frac{i}{\hbar }(p_1 -
p_2 )q} dp_1 dp_2 dq}$ \\
$= \int\limits_{ - \infty }^{ + \infty } {\chi ^{ - 1} (p_1
,p';p_2 ,p'')W_ +  {}^ - (\frac{1}{2}(p_1  + p_2 ),q_1
)e^{\frac{i}{\hbar }(p_1  - p_2 )q} dp_1 dp_2 } dq$\\
$\times\int\limits_{ - \infty }^{ + \infty } {\chi ^{ - 1} (p_1
,p';p_2 ,p'')W_ -  {}^ + (\frac{1}{2}(p_1  + p_2 ),q_2
)e^{\frac{i}{\hbar }(p_1  - p_2 )q} dp_1 dp_2 dq}$
\end{tabular} . \label{f54}
\end{equation}
Here we have introduced the following generalized functions:
\begin{equation}
\varepsilon ^{ - 1} (p_1 ,p';p_2 ,p'') = \sum\limits_{nm}
{\varepsilon ^{ - 1} (m,n)\varphi _m^ *  (p')\varphi _m (p_1
)\varphi _n (p'')\varphi _n^ *  (p_2 )}, \label{f55}
\end{equation}
\begin{equation}
\chi ^{ - 1} (p_1 ,p';p_2 ,p'') = \sum\limits_{nm} {\chi ^{ - 1}
(m,n)\varphi _m^ *  (p')\varphi _m (p_1 )\varphi _n (p'')\varphi
_n^
*  (p_2 )}. \label{f56}
\end{equation}

\section{Statistical properties of the Wigner function for charge-invariant observables. Free particle}
Constraint on the initial conditions of the Wigner function is
the general peculiarity of the approach described here because
equations are identical in both cases (for charge definite
states). In this Section we show how some theorems and properties
differ from their analogues in the usual WWM formalism and in
approach where the Newton - Wigner position operator is used.
\begin{flushleft}
\begin{itemize}
\item {\bf The property of normality.} Even part of  the Wigner
function (\ref{f38}) is normalized in the whole phase space, and
integral of the odd part (\ref{f39}) is zero.
\item {\bf The compatibility of the Wigner function (\ref{f38}) with
distributions in the coordinate and momentum spaces for a
charge-definite state.}
\begin{equation}
W_\alpha  {}^\alpha  (p) = \psi _\alpha ^ *  (p)\psi ^\alpha (p),
\label{f57}
\end{equation}
\begin{equation}
W_\alpha {} ^\alpha  (q) = \psi _\alpha ^ *  (q)\varepsilon
\left(i\hbar\overleftarrow {\partial } _q ,i\hbar\overrightarrow
{\partial } _q\right)\psi ^\alpha  (q), \label{f58}
\end{equation}
where $\psi _\alpha  (q)$ is the wavefunction in the
representation of the Newton - Wigner coordinate.
\item {\bf $n$-th moment of the coordinate} can be written as follows:
\begin{equation}
\left\langle {q^n } \right\rangle  = \int\limits_{ - \infty }^{ +
\infty } {\left\{ {\psi _\alpha ^ *  (p)\left[ {i\hbar \vec
\partial _p } \right]^n \psi ^\alpha  (p)\varepsilon (p,p')}
\right\}_{p' = p} dp}. \label{f59}
\end{equation}
The first moment (average coordinate) has a value similar to one
in the Newton-Wigner coordinate approach. Differences manifest
themselves in higher moments.
\item {\bf Criterion of pure state.}
For the functions $W_\alpha {}^\alpha (p,q)$ and $W_\alpha  {}^{
- \alpha } (p,q)$ to be even and odd components of the Wigner
function for charge-invariant observables, it is necessary and
sufficient that equalities (\ref{f42}), (\ref{f43}), (\ref{f44})
hold true, and the following conditions are satisfied:
\begin{equation}
\frac{{\partial ^2 }}{{\partial p_1 \partial p_2 }}\ln
\int\limits_{ - \infty }^{ + \infty } {W_\alpha  {}^\alpha
(\frac{1}{2}(p_1  + p_2 ),q)e^{\frac{i}{\hbar }(p_1  - p_2 )q} dq
=  - \frac{{c^4 p_1 p_2 }}{{E(p_1 )E(p_2 )(E(p_1 ) + E(p_2 ))^2
}}}, \label{f60}
\end{equation}
\begin{equation}
\frac{{\partial ^2 }}{{\partial p_1 \partial p_2 }}\ln
\int\limits_{ - \infty }^{ + \infty } {W_\alpha  {}^{ - \alpha }
(\frac{1}{2}(p_1  + p_2 ),q)e^{\frac{i}{\hbar }(p_1  - p_2 )q} dq
=  - \frac{{c^4 p_1 p_2 }}{{E(p_1 )E(p_2 )(E(p_1 ) - E(p_2 ))^2
}}}. \label{f61}
\end{equation}
\item {\bf Criterion of pure and mixed charge-definite state.}
\begin{equation}
\int\limits_{ - \infty }^{ + \infty } {W_\alpha {} ^\alpha
(p,q)\varepsilon ^{ - 2} \left(p + \frac{\hbar
}{2i}\overleftarrow{\partial}_q ,p + \frac{\hbar
}{2i}\overrightarrow{\partial}_q \right)W_\alpha {}^\alpha
(p,q)dpdq}\leq \frac{1}{{(2\pi \hbar )^d }}. \label{f62}
\end{equation}
For a pure state this inequality turns into an equality.
\end{itemize}
\end{flushleft}
\section{Statistical properties of the Wigner function for charge-invariant observables. Constant magnetic field}
\begin{flushleft}
\begin{itemize}
\item {\bf The property of normality.} Even part of  the Wigner function
(\ref{f45}), (\ref{f47}) is normalized in the whole phase space,
and integral of the odd part (\ref{f46}), (\ref{f48}) is zero.
\item {\bf The compatibility of the Wigner function (\ref{f45}),
(\ref{f47}) with distributions in the coordinate and momentum
spaces for a charge-definite state.}
\begin{equation}
W_\alpha  {}^\alpha  (p) = \psi _\alpha ^ *  (p)\varepsilon
\left(\hat {\overleftarrow{n}},\hat
{\overrightarrow{n}}\right)\psi ^\alpha (p),\label{f63}
\end{equation}
\begin{equation}
W_\alpha  {}^\alpha  (q) = \psi _\alpha ^ *  (q)\varepsilon
\left(\hat {\overleftarrow{n}},\hat
{\overrightarrow{n}}\right)\psi ^\alpha (q),\label{f64}
\end{equation}
Here $\psi _\alpha(q)$ , $\psi _\alpha(p)$ is the wavefunction in
representation of the non-local theory.
\item {\bf $n$-th moment of the
coordinate and momentum} can be written as follows:
\begin{equation}
\left\langle {q^s } \right\rangle  = \sum\limits_{nm} {q_{nm}^s
C_m ^ * {} _\alpha  \varepsilon _{mn} C_n {}^\alpha  },\label{f65}
\end{equation}
\begin{equation}
\left\langle {p^s } \right\rangle  = \sum\limits_{nm} {p_{nm}^s
C_m^ * {} _\alpha  \varepsilon _{mn} C_n {}^\alpha  }.\label{f66}
\end{equation}
The expressions for the first moments (average coordinate and
momentum) are not  similar to those in the non-local theory.
\item {\bf Criterion of pure state.}
For the functions $W_\alpha {}^\alpha (p,q)$ and $W_\alpha  {}^{ -
\alpha } (p,q)$ to be even and odd components of the Wigner
function for charge-invariant observables, it is necessary and
sufficient that equalities (\ref{f43}), (\ref{f44}), (\ref{f54})
hold true, and the following conditions are satisfied:
\begin{equation}
\frac{{\partial ^2 }}{{\partial p_1 \partial p_2 }}\ln
\int\limits_{ - \infty }^{ + \infty } {\varepsilon ^{ - 1}
(p',p_1 ;p'',p_2 )W_\alpha  {}^\alpha  (\frac{1}{2}(p' +
p''),q)e^{\frac{i}{\hbar }(p' - p'')q} dqdp'dp'' = 0}, \label{f67}
\end{equation}
\begin{equation}
\frac{{\partial ^2 }}{{\partial p_1 \partial p_2 }}\ln
\int\limits_{ - \infty }^{ + \infty } {\chi ^{ - 1} (p',p_1
;p'',p_2 )W_\alpha  ^{ - \alpha } (\frac{1}{2}(p' +
p''),q)e^{\frac{i}{\hbar }(p' - p'')q} dqdp'dp'' = 0}. \label{f68}
\end{equation}
\end{itemize}
\end{flushleft}
\section{Conclusions}
1. The usual (not Lorentz invariant) Weyl rule makes it possible
to introduce Wigner function that is not Lorentz invariant, but
all expected values calculated with it coincide with ones
calculated with Lorentz invariant wave function. This results in
fact that quantum mechanics in the Wigner formulation contains
with necessity a measuring device frame. 

2. Phase space for a scalar charged particle is not only limited
by three couples of the momenta and coordinates. The charge dimension
exists as well. However, in the approach presented here we
leave the operator nature of such variables without modification.
As a result, the matrix-valued Wigner function is the density
matrix in charge space with standard rules of expected values
calculation.

3. If we limit our consideration only to such elements of
dynamical algebra that do not depend on variables of the charge
space, it is possible to introduce the usual Wigner function. This
object differs from the Wigner function for non-relativistic
particle and from Wigner function in the Newton - Wigner position
operator approach as well. Under conditions when creation of
particles is impossible, the evolution equations coincide in the
both cases. Differences reveals themselves in the constraint on
possible initial conditions of the Wigner function.
\section*{Acknowledgement}
 Authors are very indebted to the Organizing
Committee of the 7-th International Conference on Squeezed States
and Uncertainty Relations for financial support to present this
work at the Conference.

\end{document}